\documentclass[english]{revtex4-1}
\usepackage[T1]{fontenc}
\usepackage[latin9]{inputenc}
\usepackage{amsmath}
\usepackage{amssymb}
\usepackage{graphicx}
\usepackage{wasysym}

\makeatletter
\usepackage{babel}

\makeatother

\usepackage{babel}
\begin{document}

\title{Spreading of relativistic probability densities and Lorentz contraction}

\author{Scott E. Hoffmann}

\address{School of Mathematics and Physics,~~\\
 The University of Queensland,~~\\
 Brisbane, QLD 4072~~\\
 Australia}
\email{scott.hoffmann@uqconnect.edu.au}

\begin{abstract}
We find the laws for the spreading of the spatial widths (parallel
and transverse to the direction of average motion) of the relativistic
position probability density for a massive, spinless particle. We
find that when the momentum width of the wavepacket is small compared
to the average momentum, there is a long time over which spreading
is minimal. This result may be useful in particle accelerator design.
We also demonstrate the Lorentz contraction of a wavepacket using
relativistic probability amplitudes.
\end{abstract}
\maketitle

\section{Introduction}

In a recent paper \cite{Hoffmann2018d} this author presented the
relativistic transformation properties of momentum/spin-component
and position/spin-component probability amplitudes for a massive particle
of general spin. The boost transformation properties of the position
operator were also derived. In this article, we find the rules for
the spreading with time of the spatial widths of the probability distribution
(parallel and transverse to the direction of average motion) of a
wavepacket Gaussian in momentum space. It will suffice to consider
the spinless case.

We also demonstrate the Lorentz contraction of the spatial width of
a spinless, massive, wavepacket (in the boost direction) under a boost
transformation.

\section{Derivation of the wavepacket spreading laws}

The normalized state vector we consider is
\begin{equation}
|\,\psi\,\rangle=\int\frac{d^{3}k}{\sqrt{\omega}}\,|\,k\,\rangle\Psi(k)=\int\frac{d^{3}k}{\sqrt{\omega}}\,|\,k\,\rangle\frac{e^{-|\boldsymbol{k}-\boldsymbol{p}|^{2}/4\sigma_{p}^{2}}}{(2\pi\sigma_{p}^{2})^{\frac{3}{4}}},\label{eq:1}
\end{equation}
where the basis vectors have the covariant normalization
\begin{equation}
\langle\,k_{1}\,|\,k_{2}\,\rangle=\omega_{1}\delta^{3}(\boldsymbol{k}_{1}-\boldsymbol{k}_{2})\label{eq:2}
\end{equation}
and each four-momentum is of the form $k^{\mu}=(\omega,\boldsymbol{k})^{\mu}$
with $k^{2}=m^{2}$ and $\omega=\sqrt{\boldsymbol{k}^{2}+m^{2}}.$
The factor of $1/\sqrt{\omega}$ in the superposition compensates
for the covariant normalization, as discussed in \cite{Hoffmann2018d}.

For calculation convenience, we will only consider the case where
the linear momentum distribution is strongly concentrated about the
average $\boldsymbol{p}$: $\sigma_{p}/|\boldsymbol{p}|\ll1.$

The time-dependent position wavefunction (in the Schrödinger picture)
is
\begin{equation}
\psi(t,\boldsymbol{x})=\int\frac{d^{3}k}{(2\pi)^{\frac{3}{2}}}\Psi(k)\,e^{+i(\boldsymbol{k}\cdot\boldsymbol{x}-\omega(\boldsymbol{k})t)}=\int\frac{d^{3}k}{(2\pi)^{\frac{3}{2}}}\frac{e^{-|\boldsymbol{k}-\boldsymbol{p}|^{2}/4\sigma_{p}^{2}}}{(2\pi\sigma_{p}^{2})^{\frac{3}{4}}}\,e^{+i(\boldsymbol{k}\cdot\boldsymbol{x}-\omega(\boldsymbol{k})t)}.\label{eq:3}
\end{equation}
We expand $\omega(\boldsymbol{k})$ around $\boldsymbol{k}=\boldsymbol{p}$
\begin{equation}
\omega(\boldsymbol{k})=\omega+\boldsymbol{\beta}\cdot(\boldsymbol{k}-\boldsymbol{p})+\frac{|\boldsymbol{k}-\boldsymbol{p}|^{2}}{2\omega}-\frac{1}{2\omega}(\boldsymbol{\beta}\cdot(\boldsymbol{k}-\boldsymbol{p}))^{2}+\dots\label{eq:4}
\end{equation}
where $\omega=\omega(\boldsymbol{p})$ and $\boldsymbol{\beta}=\boldsymbol{p}/\omega.$

A phase that varies by much less than unity across the wavefunction
peak region, $|\boldsymbol{k}-\boldsymbol{p}|\apprle\sigma_{p},$
will contribute negligibly to the integral. We choose to consider
only times $t$ with $|t|\ll T,$ with
\begin{equation}
\frac{\sigma_{p}^{2}}{\omega}T\sim1,\quad\mathrm{or}\quad\beta\,T\sim\frac{p}{\sigma_{p}}\sigma_{x},\label{eq:5}
\end{equation}
where $\sigma_{x}\sigma_{p}=1/2$ and $\sigma_{x}$ is the spatial
width at $t=0$ for this minimal wavepacket. With this choice, the
second order terms in Eq. (\ref{eq:4}) will be relevant, but higher
order terms can be ignored.

In the integral, Eq. (\ref{eq:3}), we change variables to $\boldsymbol{\rho}=\boldsymbol{k}-\boldsymbol{p}$
and take components $\rho_{\parallel}=\boldsymbol{\rho}\cdot\hat{\boldsymbol{\beta}}$
and $\boldsymbol{\rho}_{\perp}=\boldsymbol{\rho}-\rho_{\parallel}\hat{\boldsymbol{\beta}}.$
The measure becomes
\begin{equation}
\int d^{3}k=\int d^{3}\rho=\int d^{2}\rho_{\perp}\int_{-\infty}^{\infty}d\rho_{\parallel}.\label{eq:6}
\end{equation}
We also define components of $\boldsymbol{x}$: $x_{\parallel}=\boldsymbol{x}\cdot\hat{\boldsymbol{\beta}}$
and $\boldsymbol{x}_{\perp}=\boldsymbol{x}-x_{\parallel}\hat{\boldsymbol{\beta}}.$
Then the exponent becomes
\begin{equation}
-\frac{|\boldsymbol{k}-\boldsymbol{p}|^{2}}{4\sigma_{p}^{2}}+i(\boldsymbol{k}\cdot\boldsymbol{x}-\omega(\boldsymbol{k})t)\cong-\frac{|\boldsymbol{\rho}_{\perp}|^{2}}{4\sigma_{p}^{2}}-\frac{\rho_{\parallel}^{2}}{4\sigma_{p}^{2}}+i\rho_{\parallel}(x_{\parallel}-\beta t)-i\frac{\rho_{\parallel}^{2}t}{\gamma^{2}2\omega}+i\boldsymbol{\rho}_{\perp}\cdot\boldsymbol{x}_{\perp}-i\frac{|\boldsymbol{\rho}_{\perp}|^{2}t}{2\omega},\label{eq:7}
\end{equation}
where $\gamma=1/\sqrt{1-\beta^{2}}$ and we have ignored terms that
contribute only to a global phase factor.

We evaluate the integrals using \cite{Gradsteyn1980} (3.323.2) in
the form
\begin{equation}
\int_{-\infty}^{\infty}dz\,e^{-z^{2}/\sigma^{2}}e^{i\xi\,z/\sigma}e^{-i\tau\,z^{2}/\sigma^{2}}=\left(\frac{\pi\sigma^{2}}{1+i\tau}\right)^{\frac{1}{2}}e^{-\xi^{2}/4(1+i\tau)}.\label{eq:8}
\end{equation}
The factors we need are
\begin{equation}
e^{-\xi^{2}/4(1+i\tau)}=e^{-\xi^{2}/4(1+\tau^{2})}e^{i\xi^{2}\,\tau/4(1+\tau^{2})},\label{eq:9}
\end{equation}
with the phase factors not of interest. So we find a factor of the
form $\exp(f(\boldsymbol{x}))$ with
\begin{align}
f(\boldsymbol{x}) & =\frac{(x_{\parallel}-\beta t)^{2}4\sigma_{p}^{2}}{4\{1+(4\sigma_{p}^{2})^{2}t^{2}/\gamma^{4}4\omega^{2}\}}+\frac{|\boldsymbol{x}_{\perp}|^{2}4\sigma_{p}^{2}}{4\{1+(4\sigma_{p}^{2})^{2}t^{2}/4\omega^{2}\}}\nonumber \\
 & =\frac{(x_{\parallel}-\beta t)^{2}}{4\{\sigma_{x}^{2}+[\frac{1}{\gamma^{2}}\frac{\sigma_{p}}{|\boldsymbol{p}|}\beta t]^{2})\}}+\frac{|\boldsymbol{x}_{\perp}|^{2}}{4\{\sigma_{x}^{2}+[\frac{\sigma_{p}}{|\boldsymbol{p}|}\beta t]^{2}\}}.\label{eq:10}
\end{align}
Completing the remainder of the calculation gives the correct normalization
factors for these distributions, and will not be shown here. When
we find the position probability density, $f(\boldsymbol{x})$ is
just multiplied by 2.

So the spreading laws are
\begin{align}
\sigma_{x\parallel}(t) & =\sqrt{\sigma_{x}^{2}+\left[\frac{1}{\gamma^{2}}\frac{\sigma_{p}}{|\boldsymbol{p}|}\beta t\right]^{2}},\nonumber \\
\sigma_{x\perp}(t) & =\sqrt{\sigma_{x}^{2}+\left[\frac{\sigma_{p}}{|\boldsymbol{p}|}\beta t\right]^{2}}\quad\mathrm{for}\ \beta t\ll\frac{p}{\sigma_{p}}\sigma_{x},\label{eq:11}
\end{align}
where the variances are defined by
\begin{align}
\sigma_{x\parallel}^{2} & =\langle\,\boldsymbol{x}_{\parallel}^{2}\,\rangle-\langle\,\boldsymbol{x}_{\parallel}\,\rangle^{2},\nonumber \\
\sigma_{x\perp}^{2} & =\frac{1}{2}\{\langle\,\boldsymbol{x}_{\perp}^{2}\,\rangle-\langle\,\boldsymbol{x}_{\perp}\,\rangle^{2}\}.\label{eq:11.1}
\end{align}
We see that the spreading is not, in general, spherically symmetric
and that the long time spreading rate in the direction of average
motion is suppressed by the $1/\gamma^{2}$ factor.

The form of these spreading laws is shown in Figure 1 for $\sigma_{p}/|\boldsymbol{p}|=0.01$
and $\gamma=2.$ We see that if it was required to have negligible
spreading over the course of a scattering experiment, it would suffice
to choose the initial and final average positions of the wavepackets
so that the propagation time, $T,$ satisfied $\beta T=\sigma_{x}/\sqrt{\epsilon},$
with $\epsilon=\sigma_{p}/|\boldsymbol{p}|.$ This conclusion was
reached in \cite{Hoffmann2017a}.

\begin{figure}
\begin{centering}
\includegraphics[width=12cm]{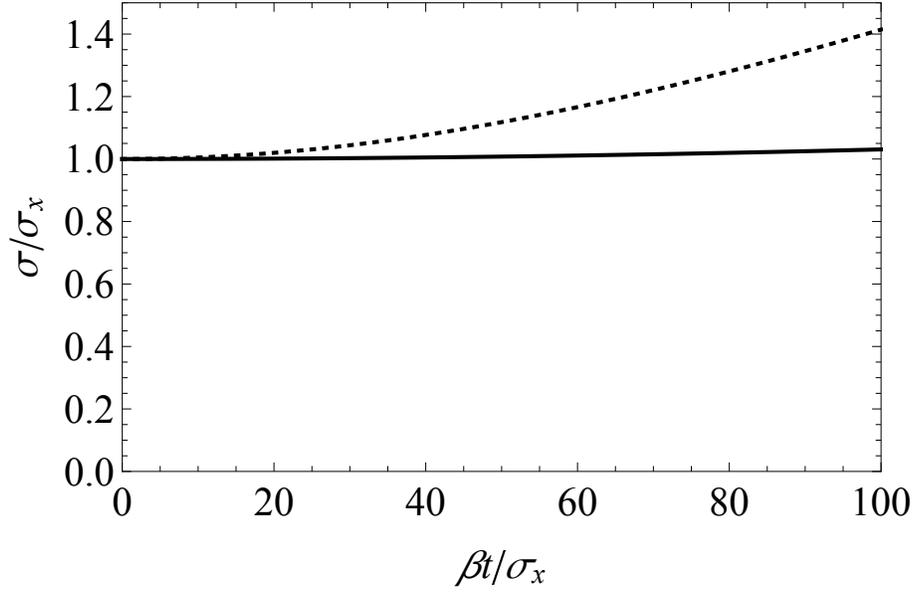}
\par\end{centering}
\caption{The scaled widths $\sigma_{x\parallel}/\sigma_{x}$ (dotted) and $\sigma_{x\perp}/\sigma_{x}$
(solid) as functions of $\beta t/\sigma_{x}$ for $\sigma_{p}/|\boldsymbol{p}|=0.01$
and $\gamma=2.$}

\end{figure}

\section{Lorentz contraction}

We consider the state vector, with $\omega=\sqrt{\boldsymbol{p}^{2}+m^{2}},$
\begin{equation}
|\,\psi\,\rangle=\int\frac{d^{3}p}{\sqrt{\omega}}\,|\,p\,\rangle\Psi(p)=\int\frac{d^{3}p}{\sqrt{\omega}}\,|\,p\,\rangle\frac{e^{-|\boldsymbol{p}|^{2}/4\sigma_{p}^{2}}}{(2\pi\sigma_{p}^{2})^{\frac{3}{4}}},\label{eq:13}
\end{equation}
representing a spinless particle with vanishing average momentum.
The position probability amplitude at $t=0$ is (up to an irrelevant
phase factor)
\begin{equation}
\psi(0,\boldsymbol{x})=\frac{e^{-|\boldsymbol{x}|^{2}/4\sigma_{x}^{2}}}{(2\pi\sigma_{x}^{2})^{\frac{3}{4}}},\label{eq:14}
\end{equation}
with $\sigma_{x}\sigma_{p}=1/2.$ So the wavepacket is minimal and
localized around the origin with a spatial width $\sigma_{x}$ at
this time.

Using the transformation results from \cite{Hoffmann2018d}, a boost
by velocity $\boldsymbol{\beta}_{0}$ produces the momentum wavefunction
\begin{equation}
\Psi^{\prime}(p)=\sqrt{\gamma_{0}(1-\boldsymbol{\beta}_{0}\cdot\boldsymbol{\beta})}\,\Psi(\Lambda^{-1}p),\label{eq:15}
\end{equation}
with $\gamma_{0}=1/\sqrt{1-\beta_{0}^{2}}$ and $\boldsymbol{\beta}=\boldsymbol{p}/\omega.$
With $p^{\prime}=\Lambda^{-1}p,$ we have
\begin{equation}
\boldsymbol{p}^{\prime}=\boldsymbol{p}_{\perp}+\gamma_{0}(\boldsymbol{p}_{\parallel}-\boldsymbol{\beta}_{0}\omega),\label{eq:16}
\end{equation}
where we have separated the momentum into parts parallel ($\boldsymbol{p}_{\parallel}$)
and perpendicular ($\boldsymbol{p}_{\perp}$) to the boost velocity.
So the exponent contains a factor
\begin{align}
|\boldsymbol{p}^{\prime}|^{2} & =|\boldsymbol{p}_{\perp}|^{2}+\gamma_{0}^{2}|\boldsymbol{p}_{\parallel}-\boldsymbol{\beta}_{0}\omega|^{2}.\label{eq:17}
\end{align}
We find that this vanishes where
\begin{equation}
\boldsymbol{p}_{\perp}=0\quad\mathrm{and}\quad\boldsymbol{p}_{\parallel}=m\gamma_{0}\boldsymbol{\beta}_{0},\label{eq:18}
\end{equation}
so the modulus-squared of the wavefunction will have its peak there.

We expand Eq. (\ref{eq:17}) in powers of $\boldsymbol{p}_{\perp}$
and $\boldsymbol{\rho}_{\parallel}=\boldsymbol{p}_{\parallel}-m\gamma_{0}\boldsymbol{\beta}_{0},$
to find
\begin{align}
|\boldsymbol{p}^{\prime}|^{2} & \cong|\boldsymbol{p}_{\perp}|^{2}+\frac{1}{\gamma_{0}^{2}}|\boldsymbol{p}_{\parallel}-m\gamma_{0}\boldsymbol{\beta}_{0}|^{2}.\label{eq:19}
\end{align}
The correction terms are of third order in the expansion quantities.
Note that the width in momentum in the boost direction will be enlarged
by the gamma factor. We choose the momentum width, $\sigma_{p},$
so that the boosted wavefunction will be narrow in momentum:
\begin{equation}
\frac{\gamma_{0}\sigma_{p}}{m\gamma_{0}\beta_{0}}=\frac{\sigma_{p}}{m\beta_{0}}\ll1.\label{eq:20}
\end{equation}
Then the third order terms we neglected in Eq. (\ref{eq:19}) will
produce higher powers of this small ratio, so are justifiably negligible.

Also we take $\boldsymbol{\beta}\rightarrow\boldsymbol{\beta}_{0},$
its peak value, in the slowly varying factor in Eq. (\ref{eq:15}),
giving
\begin{equation}
\sqrt{\gamma_{0}(1-\boldsymbol{\beta}_{0}\cdot\boldsymbol{\beta})}\rightarrow\frac{1}{\sqrt{\gamma_{0}}}.\label{eq:21}
\end{equation}

The position wavefunction at $t=0$ is then, defining $\boldsymbol{x}_{\parallel}$
and $\boldsymbol{x}_{\perp}$ as components of the position parallel
and perpendicular to $\boldsymbol{\beta}_{0},$ respectively,
\begin{equation}
\psi^{\prime}(0,\boldsymbol{x})=\int\frac{d^{3}p}{(2\pi)^{\frac{3}{2}}}e^{i(\boldsymbol{p}_{\perp}\cdot\boldsymbol{x}_{\perp}+\boldsymbol{p}_{\parallel}\cdot\boldsymbol{x}_{\parallel})}\,\frac{1}{\gamma_{0}^{\frac{1}{2}}}\,\frac{e^{-|\boldsymbol{p}_{\perp}|^{2}/4\sigma_{p}^{2}}e^{-|\boldsymbol{p}_{\parallel}-m\gamma_{0}\boldsymbol{\beta}_{0}|^{2}/4\gamma_{0}^{2}\sigma_{p}^{2}}}{(2\pi\sigma_{p}^{2})^{\frac{3}{4}}}.\label{eq:22}
\end{equation}
Evaluating the integrals using Eq. (\ref{eq:8}), as we did in the
previous section, gives
\begin{equation}
\psi^{\prime}(0,\boldsymbol{x})\propto\exp(-|\boldsymbol{x}_{\perp}|^{2}/4\sigma_{x}^{2})\exp(-\gamma_{0}^{2}|\boldsymbol{x}_{\parallel}|^{2}/4\sigma_{x}^{2}).\label{eq:23}
\end{equation}
The normalization factors are not shown, but are found to take their
correct values. Thus we see the Lorentz contraction of the spatial
width in the boost direction,
\begin{equation}
\sigma_{x\parallel}\rightarrow\frac{\sigma_{x}}{\gamma_{0}},\label{eq:24}
\end{equation}
with $\sigma_{x\parallel}$ defined as in Eq. (\ref{eq:11.1}).

It has been argued \cite{Itzykson1980} that to localize an electron
in an arbitrarily small volume would require a large amount of energy,
and that pair creation would be the inevitable result. This would
be the case if high energy photons were used to perform the localization.
We see here that the simplest way to localize a particle in an arbitrarily
small volume is to observe it from a boosted frame. The observer in
the rest frame could confirm that no pair creation was taking place.

\section{Conclusions}

Wavepackets that are narrow in momentum have a region large compared
to the minimum width, $\sigma_{x},$ over which wavepacket spreading
is negligible. This is important in modelling scattering experiments,
where we want wavepacket spreading to be minimal over the course of
the experiment. This issue was discussed by this author, in the nonrelativistic
case, for Coulomb scattering \cite{Hoffmann2017a}.

It is anticipated that the results obtained here would be useful in
accelerator design.

As a test of the physical relevance of the transformation formulae
for probability amplitudes derived in \cite{Hoffmann2018d}, we demonstrated
the Lorentz contraction, in the boost direction, of an example wavepacket.

\bibliographystyle{apsrev4-1}

\end{document}